\documentclass[english,reprint,tightenlines,eqsecnum,aps,prd,nofootinbib]{revtex4}
\usepackage[latin9]{inputenc}
\usepackage{color}
\usepackage{float}
\usepackage{amsmath}
\usepackage{amssymb}
\usepackage{graphicx}
\usepackage{esint}

\makeatletter

\newcommand{\lyxdot}{.}

\@ifundefined{textcolor}{}
{%
 \definecolor{BLACK}{gray}{0}
 \definecolor{WHITE}{gray}{1}
 \definecolor{RED}{rgb}{1,0,0}
 \definecolor{GREEN}{rgb}{0,1,0}
 \definecolor{BLUE}{rgb}{0,0,1}
 \definecolor{CYAN}{cmyk}{1,0,0,0}
 \definecolor{MAGENTA}{cmyk}{0,1,0,0}
 \definecolor{YELLOW}{cmyk}{0,0,1,0}
}

\@ifundefined{showcaptionsetup}{}{%
 \PassOptionsToPackage{caption=false}{subfig}}
\usepackage{subfig}
\makeatother

\usepackage{babel}
\begin{document}

\title{Curing the Self-Force Runaway Problem in Finite-Difference Integration}

\author{Assaf Lanir, Amos Ori and Orr Sela}

\address{Department of physics, Technion-Israel Institute of Technology, Haifa
32000, Israel}
\begin{abstract}
The electromagnetic self-force equation of motion is known to be afflicted
by the so-called \emph{runaway problem}. A similar problem arises
in the semiclassical Einstein's field equation and plagues the self-consistent
semiclassical evolution of spacetime. Motivated to overcome the latter
challenge, we first address the former (which is conceptually simpler),
and present a pragmatic finite-difference method designed to numerically
integrate the self-force equation of motion while curing the runaway
problem. We restrict our attention here to a charged point-like mass
in a one-dimensional motion, under a prescribed time-dependent external
force $F_{ext}(t)$. We demonstrate the implementation of our method
using two different examples of external force: a Gaussian and a Sin\textasciicircum{}4
function. In each of these examples we compare our numerical results
with those obtained by two other methods (a ``Dirac-type'' solution
and a ``reduction-of-order'' solution). Both external-force examples
demonstrate a complete suppression of the undesired runaway mode,
along with an accurate account of the radiation-reaction effect at
the physically relevant time scale \textemdash{} thereby illustrating
the effectiveness of our method in curing the self-force runaway problem. 
\end{abstract}
\maketitle

\section{Introduction}

This paper deals with the effect of radiation reaction on the motion
of a charged point-like mass. We develop here a pragmatic method to
numerically integrate the self-force equation of motion while overcoming
the runaway problem. Although, the main motivation for this investigation
emerges from an analogous (but considerably more complicated) problem
in semiclassical gravity: The self-consistent evolution of spacetime
geometry in response to the renormalized stress-energy tensor of quantum
fields living in spacetime. In this semiclassical problem too, one
encounters a runaway problem \cite{Wald and Flanagan,Simon 90,Simon 91,Simon 92,Stelle 77,Horowitz =000026 Wald 77};
and the technique developed here (in the charged particle context)
will potentially be helpful for addressing the analogous semiclassical
problem.

Let us first discuss the runaway problem in the case of a charged
object: In a Newtonian framework, when a charged point-like mass is
accelerated by some external force $\mathbf{F}_{ext}(t)$, it experiences
a \emph{radiation-reaction force} (or \emph{self force}), also known
as the Abraham-Lorentz force \footnote{This Newtonian expression has a straightforward relativistic extension,
known as the Abraham-Lorentz-Dirac force \cite{Dirac,Bhabha,Landau =000026 Lifshitz}
(in which a term quadratic in the acceleration is also added). The
extension of the analysis below to the relativistic case is straightforward,
but not needed here. }. It is given by \cite{Landau =000026 Lifshitz}
\begin{equation}
\mathbf{F}_{rad}=\frac{2q^{2}}{3c^{3}}\,\frac{d\mathbf{a}}{dt},\label{eq:AL-force}
\end{equation}
where $\mathbf{a}$ is the acceleration 3-vector, $q$ is the charge
and $c$ is the speed of light. \footnote{Throughout this paper we use Gaussian-cgs units.}
The motion of such a charged object therefore satisfies the equation
of motion (EOM) 
\begin{equation}
\mathbf{a}(t)=\frac{\mathbf{F}_{ext}(t)}{m}+\theta\,\frac{d\mathbf{a}(t)}{dt}\,,\label{eq:rad react intro}
\end{equation}
where $m$ is the mass and $\theta\equiv2q^{2}/3mc^{3}$. Note that
(apart from a factor $2/3$) $\theta$ is the light-travel time across
the object's ``classical radius'' $r_{c}\sim q^{2}/mc^{2}$. The
latter is the radius at which the electrostatic enrergy would become
comparable to the object's rest mass $mc^{2}$ (should the object's
size be smaller than $r_{c}$).

Despite of some potentially-confusing aspects it entails, the radiation-reaction
force (\ref{eq:AL-force}) is in itself a well-established physical
phenomenon. It acts on charged elementary particles as well as on
macroscopic charged objects \footnote{This applies as long as the object's size is small compared to the
typical wavelength of the electromagnetic waves it produces. (For
example, it would be a perfectly valid approximation for a slightly-charged
satellite in orbit around earth.) }. Yet the resultant EOM (\ref{eq:rad react intro}) suffers from an
interesting physical-mathematical pathology, known as the \emph{runaway
problem}, which we now briefly discuss.

In typical physical situations the pre-factor $\theta$ is tiny, therefore
from the physical viewpoint one expects the radiation-reaction phenomenon
to only have a small perturbative effect on the object's motion. However,
from the pure mathematical viewpoint the status of the radiation-reaction
term looks very different: Regardless of how small $\theta$ is, the
term $\theta\,d\mathbf{a}/dt$ (for any $\theta\ne$0) always constitutes
the \emph{principal part} of the ordinary differential equation (ODE)
(\ref{eq:rad react intro}). Even with the absence of any external
force, the general solution of Eq. (\ref{eq:rad react intro}) describes
an exponential growth, $\mathbf{a}(t)=\mathbf{a}_{0}\exp(t/\theta)$,
where $\mathbf{a}_{0}$ is an arbitrary vector. The presence of an
external force $\mathbf{F}_{ext}$ (assumed here to depend on $t$
only) does not change this situation much, now the general solution
is 
\begin{equation}
\mathbf{a}(t)=\mathbf{a}_{spec}(t)+\mathbf{a}_{0}\exp(t/\theta)\,,\label{eq:Exponential}
\end{equation}
where $\mathbf{a}_{spec}(t)$ is some specific solution. For a generic
solution (generic $\mathbf{a}_{0}$) the acceleration grows exponentially
at large $t$, with a time scale $\theta$. 

The absurd nature of this exact mathematical solution is made especially
clear when considering the limit of vanishing $\theta$. From the
physical viewpoint we would certainly expect the radiation-reaction
effect to vanish at this limit; but the exact mathematical solution
shows a very different picture: The rate of exponential blow-up actually
\emph{speeds up} as $\theta$ decreases, and diverges as $\theta\to0$.
This exponential acceleration of a charged object, and especially
its $\propto\theta^{-1}$ divergence, is the essence of the runaway
problem. 

For illustration, we mention here the value of $\theta$ for two very
different physical objects: (i) The electron has $\theta\simeq6.3\cdot10^{-24}$
sec. (ii) The sun is claimed to be charged \cite{Sun} by $q\approx77$
Coulombs, and correspondingly $\theta\simeq6.6\cdot10^{-43}$ sec.
Certainly such an exponentially-growing acceleration of the sun is
strikingly inconsistent with our daily astronomical observations. 

The aforementioned runaway pathology brings about two different types
of challenges: The first is how to reconcile, at the level of principle,
the weird mathematical properties of Eq. (\ref{eq:rad react intro})
with our basic physical expectations (and our daily experience!) concerning
the motion of charged objects. The second one is more pragmatic: How
to produce the physically-valid and meaningful solutions for $\mathbf{a}(t)$
out of this mathematically-problematic EOM\textcolor{red}{{} }\eqref{eq:rad react intro}.
Whereas the main objective of this paper is the second challenge,
we first briefly discuss the first one. 

The expression (\ref{eq:AL-force}) for the radiation-reaction force
may be derived in various methods. For the sake of the present discussion
we find it useful to consider the extended-object approach \cite{Extended},
which is carried entirely in the framework of classical electrodynamics.
In this approach, one considers a charged object of typical size $L$
and computes the overall (classical) electromagnetic force acting
on this object, by appropriately integrating the mutual forces acting
on the object's various volume elements. Then one takes the limit
where the object's size L shrinks to zero. This yields the relativistic
Abraham-Lorentz-Dirac equation \textemdash{} which in the Newtonian
framework reduces to Eq. (\ref{eq:AL-force}). This method reveals
two subtleties that must be kept in mind: (a) The expression (\ref{eq:AL-force})
(like its relativistic counterpart) is only obtained at the $L\to0$
limit. For any finite $L$ there are small finite-size corrections.
The typical magnitude of these corrections is $\sim t_{L}/\,t_{orb}$
where $t_{L}\equiv L/c$ is the light-travel time across the object.
Here $t_{orb}$ denotes the typical dynamical time scale characterizing
the object's motion (namely the typical time scale for changing the
velocity, acceleration, etc.). For a wide range of object's parameters
$L,m,q$, and external forces $\mathbf{F}_{ext}$, these finite-size
corrections are negligible (for a physically valid orbit), but not
strictly vanishing. (b) For orbits in which $t_{orb}$ is small compared
to $t_{L}$, the expression (\ref{eq:AL-force}) becomes invalid (as
in fact implied from point a). 

The reason for the failure of the naive mathematical solutions (\ref{eq:Exponential})
of the EOM (\ref{eq:rad react intro}) now becomes clear: In such
exponentially-accelerating solutions $t_{orb}$ becomes $\theta$,
which always fails to be $\gg t_{L}$. \footnote{The situation $\theta\gg t_{L}$ would require the object's size $L$
to be $\ll r_{c}$, which would in turn imply that the object's bare
mass is negative \textemdash{} which is presumably impossible for
a classical object. (It should be emphasized that our mass parameter
$m$ is the object's \emph{observed mass}, and the bare mass is obtained
by subtracting the object's electrostatic energy.)} For example, for the charged sun $t_{L}/\theta$ is $\approx7\cdot10^{42}$
rather than $\ll1$ (for the electron it is of order unity if we take
$L$ to be its classical radius). Hence the Abraham-Lorentz force
formula is totally invalid for such solutions \textemdash{} explaining
the physical irrelevance of the runaway solutions. 

The above discussion clarifies the nature of the solutions of the
EOM (\ref{eq:rad react intro}) that would have physical relevance:
First, only solutions with sufficiently large $t_{orb}$ (compared
to $t_{L}$) would be relevant. Second, we should \emph{not} expect
the actual physical orbit to be an exact\textcolor{red}{{} }solution
of the mathematical EOM: We do expect small deviations (in the actual
radiation-reaction force) of typical order $t_{L}/t_{orb}\ll1$. Third,
in principle the physical solutions should be causal. Namely, the
value of $\mathbf{F}_{ext}$ at a given moment should not affect the
orbit at earlier times.

The case that mostly concerns us in this paper is that of weakly charged
objects, yielding small $\theta$ (this is the case in which the runaway
problem becomes most acute). As it turns out, in this case there is
a class of physically-meaningful solutions that naturally suggests
itself, which satisfies all the above physical requirements: These
are the solutions in which the self force only yields a tiny correction
to the object's orbit. Namely, the deviation of the actual $\mathbf{a}(t)$
from $\mathbf{F}_{ext}/m$ is small. In these solutions the orbit's
time scale $t_{orb}$ is just the one set by the external force, which
we denote by $\tau$ {[}namely, the characteristic time scale for
significant changes in $\mathbf{F}_{ext}(t)${]}. The assumption of
``small $\theta$'' can now be formulated more explicitly: We shall
hereafter assume that $\theta\ll\tau$. This separation of time scales
turns out to be essential in the treatment of the runaway problem. 

Let us briefly summarize the discussion so far, concerning the motion
under weak radiation reaction ($\theta\ll\tau$): The generic, exact,
mathematical solution of the EOM (\ref{eq:rad react intro}) runs-away
exponentially, as seen in Eq. (\ref{eq:Exponential}), with an exponential
time scale $t_{orb}=\theta$. However, since this time scale fails
to be $\gg t_{L}$, these runaway solutions are physically invalid
(because the Abraham-Lorentz formula then fails to describe the correct
radiation-reaction force). But there is another class of (approximate)
non-runaway solutions. In these solutions $t_{orb}\sim\tau$ (which
is presumably $\gg t_{L}$), hence the Abraham-Lorentz expression
(\ref{eq:AL-force}) is valid. These are the only (approximate) solutions
of Eq. (\ref{eq:rad react intro}) which are of physical relevance.
\footnote{There also exist the class of ``Dirac-type'' (described below),
which are exact solutions of the EOM (\ref{eq:rad react intro}) with
no runaway. But these solutions are a-causal. } We emphasize again that from the physical viewpoint there is no reason
to insist on exact solutions to the EOM (\ref{eq:rad react intro}),
due to the inherent finite-size error in the radiation-reaction term.
This error is of typical order $t_{L}/t_{orb}$, which in the physically-relevant
solutions becomes $t_{L}/\tau\ll1$ (and is always $\gtrsim\theta/\tau$). 

After we clarified the nature of the desired radiation-reaction solutions,
we come to the next challenge: How to construct these physically relevant
solutions from the mathematically-problematic EOM (\ref{eq:rad react intro})?
Addressing this question is the main goal of this paper. Note that
this is not a trivial task, because the actual mathematical solutions
of this ODE diverge exponentially at large $t$ (except for a subclass
of measure zero \textemdash{} the Dirac-type solutions discussed below
\textemdash{} which however turns out to be a-causal). 

Our working assumption here is that at the end of the day it will
be necessary to integrate the EOM numerically (while overcoming the
runaway problem). Whereas for simple examples of external forces $\mathbf{F}_{ext}(t)$
an analytical solution may be possible, once $\mathbf{F}_{ext}$ is
allowed to depend on location the situation changes and an analytical
solution becomes unlikely. Furthermore, in the problem which really
motivates this investigation \textemdash{} the self-consistent semiclassical
evolution of spacetime \textemdash{} the solution of the field equation
(the semiclassical Einstein equation) ought to be numerical. 

Correspondingly, our approach will be to handle the runaway problem
while numerically integrating the EOM (\ref{eq:rad react intro}),
using the finite-difference approach. Recall that the desired physical
solutions for the orbit should have a time scale $t_{orb}\sim\tau$,
which is presumably larger than the runaway time-scale $\theta$ by
many orders of magnitude. Furthermore, we have no intention to numerically
reproduce the (unphysical) runaway dynamics at the timescale $\theta$:
Instead, we would like our numerical integration procedure to simply
kill this undesired runaway. As we shall show below, this goal can
be naturally achieved if we take the numerical time-step $\delta t$
to be in the range $\theta\ll\delta t\ll\tau$. \footnote{The condition $\delta t\ll\tau$ is necessary for achieving an accurate
numerical solution (at the relevant time-scale $\tau$). Eventually
it turns out that the other condition $\delta t\gg\theta$ is not
really necessary, it may be replaced by the weaker one $\delta t$
$>2\theta$ (see Sec. \ref{sec:Discretizing}). \label{fn:The-condition}} We shall design our finite-difference scheme so as to entirely prevent
the growth of the undesired runaway mode, while accurately accounting
for the radiation-reaction effect at the physically relevant time
scale $\tau$. 

As was already mentioned above, the physical problem which really
motivates this work is the self-consistent semiclassical evolution
of e.g. an evaporating macroscopic black hole. In this case spacetime
evolution is governed by the semiclassical Einstein equation 
\begin{equation}
G_{\mu\nu}\left(g_{\alpha\beta}\right)=8\pi G\left\langle \hat{T}_{\mu\nu}\left(g_{\alpha\beta}\right)\right\rangle ,\label{eq:semiclassical Einstein}
\end{equation}
where the unknown $g_{\alpha\beta}$ is the spacetime metric, $G_{\mu\nu}$
is the Einstein tensor, $G$ is Newton's gravitational constant, and
$<\hat{T}_{\mu\nu}>$ is the (expectation value of the) renormalized
stress-energy tensor. This equation describes the back-reaction of
a quantum field on spacetime. Recall that the Einstein tensor $G_{\mu\nu}$
is a second-order differential operator acting on $g_{\alpha\beta}$.
The right-hand side, being proportional to $\hbar$, is supposed to
be a small correction term; however, it contains higher-order derivatives
of the metric up to forth order, \footnote{The source term $<\hat{T}_{\mu\nu}>$ is basically made of the contributions
of the various modes of the quantum field. However, the sum/integral
over the modes diverges, and requires renormalization. In curved spacetime
the renormalization is usually done by point-splitting, and it involves
the subtraction of certain counter-terms \cite{Christensen a,Christensen b}
made of the Riemann (or Ricci) tensor and its derivatives up to second
order. This amounts to fourth-order derivatives of the basic unknown
$g_{\alpha\beta}$. } formally leading to a runaway instability in the semiclassical metric
evolution. This runaway problem is regarded \cite{Wald's red book}
as one of the most serious obstacles in the attempt to obtain the
physically relevant self-consistent solutions of the semiclassical
Einstein equation (\ref{eq:semiclassical Einstein}). 

As was pointed out in Ref. \cite{Wald and Flanagan}, there is a remarkable
analogy between the two problems: In both cases, we originally have
a second-order differential equation {[}for the particle's orbit $\mathbf{r}(t)$
or for the spacetime metric $g_{\alpha\beta}(x)${]}; and there is
a back-reaction term which is multiplied by a small pre-factor, yet
it includes higher-order derivatives. In both problems, therefore,
there is a runaway catastrophe: The general mathematical solution
of the back-reaction equation blows up exponentially with a ridiculously
small time scale (which becomes shorter as the pre-factor of the back-reaction
term decreases). In both cases the relevant physical solutions are
those with dynamical time scales much longer than the runaway timescale.
Owing to these similarities, we anticipate that the finite-difference
method developed here will successfully cure the runaway problem in
the semiclassical context as well. 

In the charged-object case, in order to obtain physically meaningful
solutions to Eq. \eqref{eq:rad react intro}, two different approaches
have been introduced previously. Both approaches are further described
and implemented in Sec. \ref{sec:Numerical-implementations} below,
in order to compare them with our new finite-difference method. The
first, introduced by Dirac \cite{Dirac}, assumes that the external
force should vanish at late time. Correspondingly it considers only
the solutions to Eq. (\ref{eq:rad react intro}) for which $\mathbf{a}(t)$
vanishes at late time. This prescription inherently violates causality,
but nevertheless the time scale of the violation is minuscule, it
coincides with the runaway timescale $\theta$ (e.g. for the electron
it is $\sim6\cdot10^{-24}s$). The second approach is the method of
\emph{reduction of order} \cite{Bhabha,Landau =000026 Lifshitz,Teitelboim,Ford}.
In this method, one replaces $d\mathbf{a}/dt$ in Eq. \eqref{eq:rad react intro}
with the time derivative of the external force (see Sec. \ref{sec:Numerical-implementations}).
Note that this procedure involves neglecting a term quadratic in the
small parameter $\theta/\tau$. 

Both approaches, Dirac's method and reduction of order, seem to be
inapplicable (or at least impractical) in the problem of self-consistent
semiclassical evolution of e.g. an evaporating black hole. Dirac's
method relies on requiring a trivial asymptotic behavior of the solution
at the infinite future. This seems to be inapplicable to semiclassical
evolution of black holes, due to the presence of a future singularity
inside the horizon. In addition, since we do not have at our disposal
a closed-form solution of the semiclassical Einstein equation, there
is no way to know in advance which initial conditions would lead to
a desired late-time solution (even if such a desired solution existed).
The method of reduction of order works well in the case of ODEs like
Eq. (\ref{eq:rad react intro}), in which there is a single higher-order
derivative term ($d\mathbf{a}/dt$) that needs be eliminated, which
can be directly obtained by differentiating the ``original'' EOM
$a=\mathbf{F}_{ext}/m$. This is not the case in our semiclassical
problem, mainly because the semiclassical Einstein equation (\ref{eq:semiclassical Einstein})
is a \emph{partial} differential equation (PDE), making it hard to
eliminate the various higher-order derivative terms involved in $<\hat{T}_{\mu\nu}>$.
We point out that the method of reduction of order has been applied
to several semiclassical gravitational systems (see, for example,
\cite{Bel,Simon =000026  Parker}); but to the best of our knowledge,
in all these cases either (i) the semiclassical Einstein equation
reduced to an ODE, or (ii) the quantum field was a conformal scalar
field. (In the latter case no derivatives of the Riemann tensor appear
in the counter-term. Only derivatives of the Ricci scalar and Ricci
tensor appear, which can in turn be expressed by the energy-momentum
tensor via Einstein equation.) Both simplifying factors (i) or (ii)
allow for easy reduction of order. This leaves unresolved the problem
of doing reduction of order in the situation of e.g. black-hole evaporation
(via quantum fields other than conformal scalar field). 

On the other hand, the finite-difference method introduced here will
hopefully be applicable to the problem of semiclassical black hole
evaporation as well. Of course this problem is technically much harder
than the electromagnetic radiation reaction problem, due to several
reasons. One obvious difference is that the semiclassical problem
involves PDEs rather than ODEs. As far as we can foresee, this difference
should not disable our finite-difference method (although it may possibly
make its application technically harder). \footnote{A much more crucial difference is that in the semiclassical problem
it is very difficult to compute the source term $<\hat{T}_{\mu\nu}>$.
In particular it requires the computation of many field's modes on
the background metric $g_{\mu\nu}$, combined with the point-splitting
\cite{Christensen a,Christensen b} regularization procedure. \label{fn:RSET}}

Throughout this paper we shall restrict our attention to one-dimensional
motion, and also to a prescribed external force $\mathbf{F}_{ext}(t)$
that depends on time only. In a forthcoming paper \cite{next paper}
we extend the method to an external force depending on space as well
as time, yielding a third-order radiation-reaction EOM in terms of
the position (rather than a first-order equation for the acceleration).
We shall show that our finite-difference method works very well in
this generalized framework as well. This will be carried out as an
intermediate stage towards the application of the procedure to a hyperbolic
PDE with an artificially-added higher-derivative term \textemdash{}
before applying it finally to the semiclassical Einstein field equation. 

The rest of this paper is organized as follows. In Sec. \ref{sec:Discretizing}
we construct the finite-difference scheme which resolves the runaway
problem. In Sec. \ref{sec:Numerical-implementations} we numerically
implement our finite-difference method to two examples of external
force, one is Gaussian and the other is of a Sin\textasciicircum{}4
form. We compare our numerical results with the corresponding Dirac
solution and the solution obtained by reduction of order, and we find
good agreement. We then summarize our results and give some concluding
remarks in Sec. \ref{sec:Conclusions}.

\section{Discretizing the equation of motion \label{sec:Discretizing}}

Our starting point is the Newtonian EOM (\ref{eq:rad react intro}),
which we take now to be one-dimensional (say in the $x$ direction):
\begin{equation}
a(t)=\frac{F_{ext}(t)}{m}+\theta\,\frac{da(t)}{dt}\,,\label{eq:physical nonhom}
\end{equation}
where, recall, $\theta\equiv2q^{2}/3mc^{3}$. We assume that the external
force $F_{ext}$ only depends on $t$, not on $x$. Recall that there
are two different time scales in this EOM: the \emph{external-force
time scale} $\tau$, and the \emph{runaway time scale} $\theta$;
and throughout this paper we are interested in the case of weak radiation
reaction, $\theta\ll\tau$ (the situation that mostly sharpens the
runaway problem.) Our goal is to design a stable numerical discretization
scheme for such an EOM, free of runaway. 

As a first step, we use the time parameter $\tau$ to transform the
variables in Eq. (\ref{eq:physical nonhom}) into dimensionless ones:
\[
t\to T\equiv t/\tau\;,\;\;\;a\to A\equiv(\tau/c)a\;,\;\;\;F_{ext}\to F\equiv(\tau/mc)F_{ext}\;.
\]
The resulting dimensionless EOM is
\begin{equation}
A=\varepsilon\dot{A}+F(T)\,,\label{eq: nonhom acceleration}
\end{equation}
where an over-dot denotes $d/dT$, and $\varepsilon\equiv\theta/\tau$.
Our assumption of weak radiation reaction now reads $\varepsilon\ll1$. 

\begin{figure}
\begin{centering}
\includegraphics[scale=0.7]{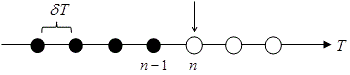}
\par\end{centering}
\caption{Illustration of the lattice points in the numerical scheme. The full
black spots mark the points $k<n$ for which $A_{k}$ has already
been computed. The empty circles denote points $k\geq n$ for which
the values of $A_{k}$ are still unknown. The vertical arrow points
at the $n$th lattice point, where the value of $A$ is being computed
at this stage. The distance between each pair of neighboring points
is the ``lattice constant'' $\delta T$. \label{fig.  lattice}}
\end{figure}

In a finite-difference numerical scheme, we treat the time axis as
a discrete set of ``lattice'' points $T_{k}$, with a fixed separation
$\delta T=T_{k+1}-T_{k}$ (see Fig. \ref{fig.  lattice}). In each
step of the numerical computation, the value of $A\left(T\right)$
is already known on all lattice points $k<n$ (for some $n$), but
$A_{n}$ (like all points $A_{k>n})$ is still unknown. The numerical
scheme should then determine $A_{n}$, expressing it in terms of (some
of) the already-known points $A_{k<n}$ (the black spots in Fig. \ref{fig.  lattice}).
We shall restrict the discussion here to the simplest example for
the discretization of a first-order ODE, which we call \emph{two-point
discretization}, that only involves the two lattice points $A_{n-1}$
and $A_{n}$. 

We need to (approximately) express the various elements of Eq. (\ref{eq: nonhom acceleration})
in terms of the two available data points $A_{n-1}$ and $A_{n}$.
There is an obvious expression for the discretized derivative $\dot{A}$:
\begin{equation}
\dot{A}(T)\rightarrow\frac{A_{n}-A_{n-1}}{\delta T}.\label{eq:a dot discretization}
\end{equation}
However, a priori there is no unique choice for the discretization
of $A(T)$. For example, one may choose to represent it by $A_{n}$,
or $A_{n-1}$, or by some weighted average of the two. Each choice
will lead to a slightly different finite-difference scheme. At this
stage we find it useful to allow for a general representation 
\begin{equation}
A(T)\rightarrow\gamma A_{n}+(1-\gamma)A_{n-1},\label{eq:a discretization}
\end{equation}
with a (real) free parameter $\gamma$. The representation of $F(T)$
is done similarly, using the same weighted average: 
\begin{equation}
F(T)\rightarrow\gamma F_{n}+(1-\gamma)F_{n-1},\label{eq:a discretization-1}
\end{equation}
where $F_{n}\equiv F\left(T_{0}+n\delta T\right)$, and $T_{0}$ is
the $T$ value corresponding to the first lattice point $n=0$. The
discretized (dimensionless) EOM then becomes
\begin{equation}
\gamma A_{n}+(1-\gamma)A_{n-1}=\epsilon\left(A_{n}-A_{n-1}\right)+\left[\gamma F_{n}+(1-\gamma)F_{n-1}\right]\,,\label{eq:discetized_full}
\end{equation}
with a free discretization parameter $\gamma$, where $\epsilon\equiv\varepsilon/\delta T$.

\subsection{Choosing the discretization parameter $\gamma$ \label{subsec:Choosing-gamma}}

As it turns out, the choice of $\gamma$ will affect the dynamical
behavior of the finite-difference scheme. We shall take advantage
of this free parameter, in order to achieve our goal of suppressing
the problematic runaway mode. 

Let $A_{phys}(T)$ denote the solution of the ODE \eqref{eq: nonhom acceleration}
which we regard as the desired physical solution (i.e. the one which
does not exhibit exponential runaway). Consider now another solution
$A(t)$ which (at some moment $T$) slightly deviates from $A_{phys}(T)$.
We denote the difference by $\Delta A(T)\equiv A(T)-A_{phys}(T)$.
This deviation satisfies the homogeneous equation 
\begin{equation}
\Delta A=\varepsilon\Delta\dot{A}\,.\label{eq:homogeneous}
\end{equation}
Mathematically, every small initial deviation $\Delta A$ from $A_{phys}(T)$
will grow exponentially in $T$: Obviously the general solution of
Eq. (\ref{eq:homogeneous}) is $\Delta A(T)=const\cdot\exp(T/\varepsilon)$.
In order for the numerical scheme to stably recover the desired solution
$A_{phys}(T)$, our finite-difference scheme must suppress this homogeneous
mode. Applying the above discretization scheme to the homogeneous
equation (\ref{eq:homogeneous}) yields 

\begin{equation}
\gamma\Delta A_{n}+(1-\gamma)\Delta A_{n-1}=\epsilon\left(\Delta A_{n}-\Delta A_{n-1}\right)\,.\label{eq:discetized_hom}
\end{equation}
Extracting $\Delta A_{n}$ we obtain 
\begin{equation}
\Delta A_{n}=p\,\Delta A_{n-1}\,,\label{eq:an simple}
\end{equation}
where 
\[
p=1-\frac{1}{\gamma-\epsilon}\,.
\]
This formula describes a geometric sequence, $\Delta A_{n}=const\cdot p^{n}$.
The qualitative nature of this sequence will depend crucially on whether
$|p|$ is greater or smaller than one: The sequence $\Delta A_{n}$
will blow up (as $n\to\infty$) in the first case and decay in the
second. We want our finite-difference scheme to converge to the desired
solution $A_{phys}(T)$, which would correspond to $\Delta A(t)\to0$
(and the faster the better). The range of $\gamma$ which satisfies
this stability condition is $\gamma>\epsilon+1/2$. 

We find it especially convenient to work with $\gamma=1$ (implying
that $A(t)$ is simply discretized as $A_{n}$). This choice of $\gamma$
corresponds to the smallest value of $|p|$ (and hence to the quickest
damping of $\Delta A_{n}$), if: (i) we consider a $\gamma$ choice
independent of $\epsilon$, and (ii) we recall that in typical situations
we shall have $\epsilon\ll1$, which (for $\gamma=1$) would yield
$p\approx-\epsilon$. 

\subsection{The final discretization scheme}

Substituting $\gamma=1$ in Eq. (\ref{eq:discetized_full}) and extracting
$A_{n}$, we obtain the discretized EOM: 
\begin{equation}
A_{n}=-\frac{\epsilon}{1-\epsilon}A_{n-1}+\frac{1}{1-\epsilon}\,F_{n}\,,\label{eq:an nonhom}
\end{equation}
where, recall, $\epsilon\equiv\varepsilon/\delta T$. The deviations
from the desired solution $A_{phys}(T)$ will thus decay according
to the geometric sequence
\begin{equation}
\Delta A_{n}=const\cdot p^{n}\;\;,\;\;\;\;\;p=-\frac{\epsilon}{1-\epsilon}\,.\label{eq:p_hom}
\end{equation}

As was already mentioned in the Introduction, in a typical situation
(with a huge separation of time scales $\theta\ll\tau$) we would
typically choose the numerical time step $\delta t$ in the range
$\theta\ll\delta t\ll\tau$ . In terms of the dimensionless parameters
this reads $\varepsilon\ll\delta T\ll1$. The condition $\delta T\ll1$
is really necessary for controlling the truncation error. The other
condition $\delta T\gg\varepsilon$ is not mandatory but merely a
matter of typicality and convenience. Note, however, that we must
always require

\begin{equation}
\delta T>2\varepsilon\label{eq:Mandatory}
\end{equation}
in order to achieve numerical stability. This directly follows from
Eq. (\ref{eq:p_hom}), which indicates that the condition for having
$|p|<1$ is $\epsilon<1/2$.

The choice of $\delta T$ in the appropriate range ($2\varepsilon<\delta T\ll1$)
guarantees that no runaway mode (and no numerical-instability mode)
will develop. The evolving numerical solution will therefore be characterized
by the external-force time scale; and since the time step is much
smaller than that scale, the truncation error will be well controlled
and the numerical solution should faithfully recover the desired physical
solution. The effectiveness of this numerical procedure will be demonstrated
in the next section. 

\subsubsection*{Initial value for $A$}

In principle, in the numerical scheme described above we need to choose
the initial value of $A$, which we denote $A_{0}$. Owing to the
quick suppression of the homogeneous mode, the resultant numerical
solution is insensitive to the choice of $A_{0}$ (except in a very
short interval at the beginning, a few times $\delta T$). We shall
hereafter adopt a standard choice: 
\begin{equation}
A_{0}=F(T=0).\label{eq:Initial_A}
\end{equation}

\section{Numerical implementations \label{sec:Numerical-implementations}}

We shall now demonstrate this finite-difference method numerically,
using two different examples of external force $F(T)$. In both examples
we shall compare our numerical result with the corresponding Dirac-type
solution and with the solution of the reduction-of-order variant of
Eq. \eqref{eq: nonhom acceleration}. As a preparatory step we shall
now develop the analytical expressions for the Dirac-type and the
reduction-of-order solutions, for a general $F(T)$. 

We begin with the general solution of Eq. \eqref{eq: nonhom acceleration},
written in the form
\[
A(T)=-\frac{1}{\varepsilon}e^{T/\varepsilon}\int_{\infty}^{T}e^{-s/\varepsilon}F(s)ds+Ce^{T/\varepsilon}\,,
\]
where $C$ is an arbitrary constant. Assuming a function $F(T)$ of
a bounded support (or an ``effectively-bounded'' support as in the
Gaussian case), at large $T$ this solution approaches $Ce^{T/\varepsilon}$.
The Dirac-type solution, which we denote $A_{D}(T)$, is the unique
solution that vanishes as $T\rightarrow\infty$ \textemdash{} namely
the $C=0$ solution:
\begin{equation}
A_{D}(T)=-\frac{1}{\varepsilon}e^{T/\varepsilon}\int_{\infty}^{T}e^{-s/\varepsilon}F(s)ds\,.\label{eq:Dirac General}
\end{equation}

Next, we apply the reduction-of-order procedure (to first order in
$\varepsilon$), and construct the corresponding analytical expression,
which we denote $A_{r}(T)$. To this end, we first differentiate Eq.
\eqref{eq: nonhom acceleration} with respect to $T$: 
\[
\dot{A}=\varepsilon\ddot{A}+\dot{F}(T)\,.
\]
 Inserting this into the right-hand side of Eq. \eqref{eq: nonhom acceleration}
we get 
\[
A(T)=\varepsilon\left[\varepsilon\ddot{A}(T)+\dot{F}(T)\right]+F(T)=\varepsilon\dot{F}(T)+F(T)+O(\varepsilon^{2})\,.
\]
 In constructing the reduction-of-order solution, we simply ignore
the $O\left(\varepsilon^{2}\right)$ term and obtain:
\begin{equation}
A_{r}(T)=F(T)+\varepsilon\dot{F}(T)\,.\label{eq:reduction}
\end{equation}
We point out that the Dirac-type solution is in principle a-causal,
as clearly indicated by the lower integration limit $s\to\infty$
in Eq. (\ref{eq:Dirac General}) (nevertheless this a-causality is
usually small, the advance time is typically of magnitude $\varepsilon$).
The reduction-of-order expression $A_{r}(T)$ does not suffer from
such violation of causality. This expression is not an exact solution
of the EOM \eqref{eq: nonhom acceleration}, but nevertheless it is
a solution to first order in $\varepsilon$. 

\subsection{Gaussian External Force}

\begin{figure}
\begin{centering}
\subfloat[]{\centering{}\includegraphics[scale=0.4]{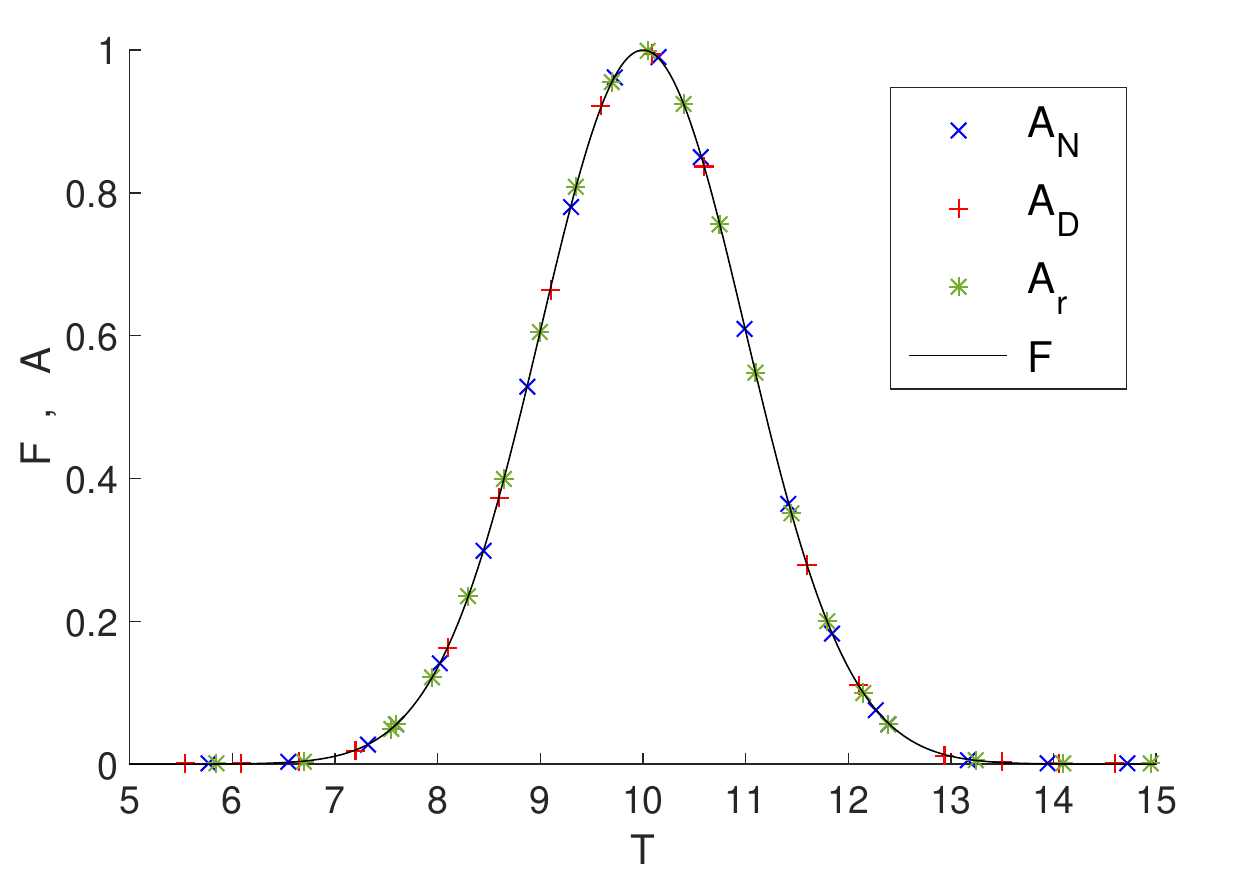}}\subfloat[]{\centering{}\includegraphics[scale=0.4]{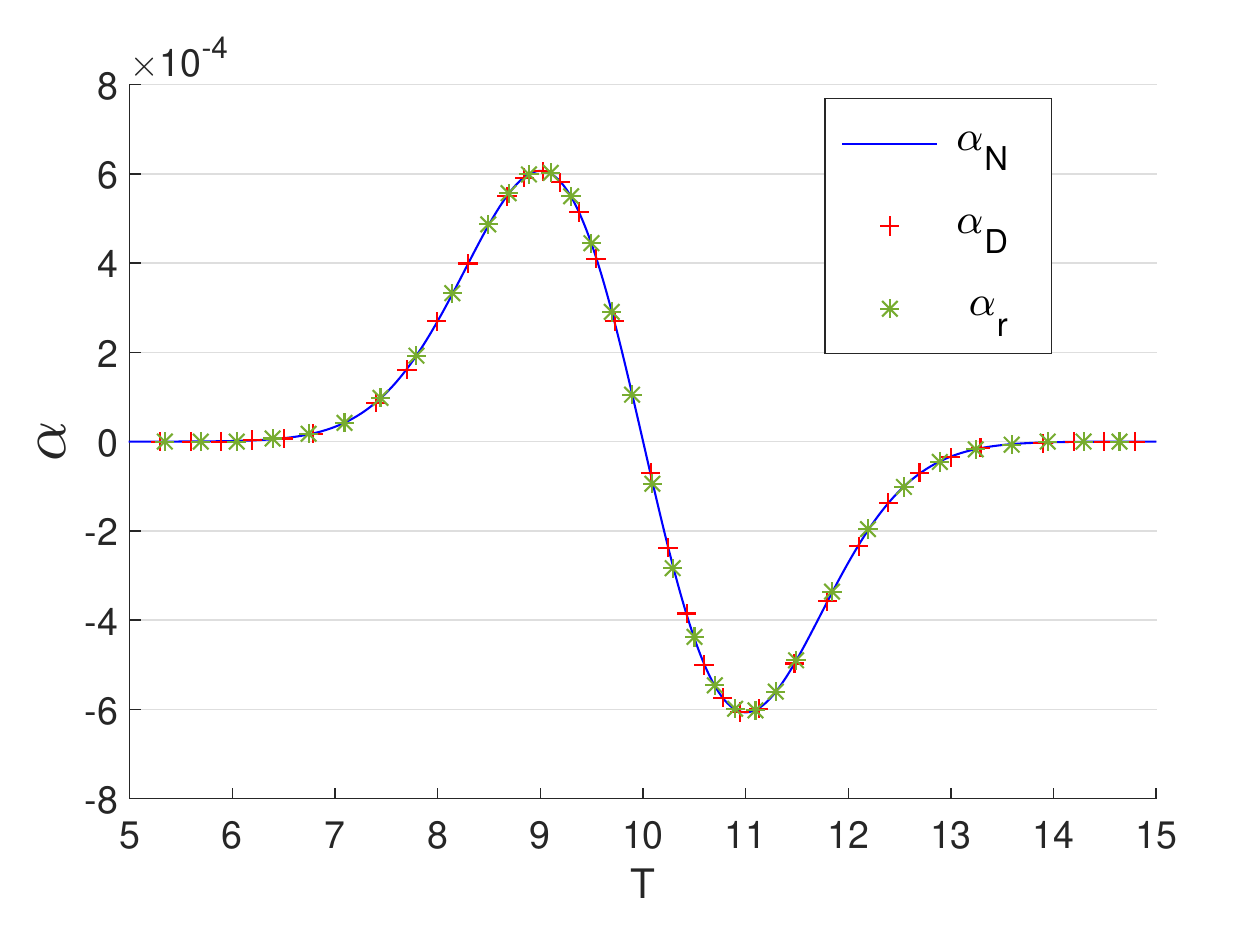}}
\par\end{centering}
\caption{Gaussian external force. Panel (a) displays the external force $F(T)$
(black solid line) given by Eq. \eqref{eq:gaussian force}, with parameters
$B=\sigma=1$ and $\mu=10$. It also displays the solution $A(T)$
of the EOM \eqref{eq: nonhom acceleration} corresponding to this
Gaussian external force. This solution comes in three variants: Our
numerically-constructed solution $A_{N}(T)$, and the two analytically-constructed
solutions $A_{D}(T)$ (Dirac-type solution) and $A_{r}(T)$ (reduction
of order), used here for comparison. These three variants of $A(T)$
are respectively represented by blue crosses, red ``plus'' symbols,
and green stars. However these three accelerations are visually indistinguishable
from the black curve of $F(T)$ (and from each other). Panel (b) displays
the differences $\alpha(T)\equiv A(T)-F(T)$ between the external
force and the acceleration(s) (which directly represent the effect
of the radiation-reaction force). There are three variants of $\alpha$,
in correspondence with the three variants of $A$: The blue solid
curve denotes $\alpha_{N}=A_{N}-F$, the red ``plus'' symbols denote
$\alpha_{D}=A_{D}-F$, and the green stars denote $\alpha_{r}=A_{r}-F$.
The numerically-constructed function $\alpha_{N}$ is visually indistinguishable
from the two analytically-constructed functions $\alpha_{D}$,$\alpha_{r}$
\textemdash{} demonstrating the accuracy of our finite-difference
scheme. In both panels the numerical time step was $\delta T=5\cdot10^{-3}$
(correspondingly the number of steps was $N=4\cdot10^{3}$). \label{fig:gaussian} }
\end{figure}

As a first example, we choose $F(T)$ to be a Gaussian, 
\begin{equation}
F(T)=B\exp\left[-\frac{(T-\mu)^{2}}{2\sigma^{2}}\right],\label{eq:gaussian force}
\end{equation}
as displayed in Fig. \ref{fig:gaussian}a. The corresponding Dirac-type
solution (\ref{eq:Dirac General}) is then 
\begin{equation}
A_{D}(T)=\frac{B}{\varepsilon}\sqrt{\frac{\pi\sigma^{2}}{2}}\exp\left(\frac{T}{\varepsilon}+\frac{\sigma^{2}/\varepsilon-2\mu}{2\varepsilon}\right)\mathrm{Erfc}\left(\frac{\sigma^{2}/\varepsilon+T-\mu}{\sqrt{2\sigma^{2}}}\right),\label{eq:dirac gauss}
\end{equation}
where $\mathrm{Erfc}$ is the complementary error function given by
\[
\mathrm{Erfc}(x)=\frac{2}{\sqrt{\pi}}\intop_{x}^{\infty}\exp\left(-s^{2}\right)ds\,.
\]
The reduction-of-order solution (\ref{eq:reduction}) now reads
\begin{equation}
A_{r}(T)=\left[1-\frac{\varepsilon}{\sigma^{2}}\left(T-\mu\right)\right]B\exp\left[-\frac{(T-\mu)^{2}}{2\sigma^{2}}\right]\,.\label{eq:reduced gauss}
\end{equation}

Specifically we choose here the parameters $B=\sigma=1$ and $\mu=10$.
For the radiation-reaction strength we choose $\varepsilon=10^{-3}$.
We then solve Eq. \eqref{eq: nonhom acceleration} numerically using
the finite-difference scheme constructed above, Eq. (\ref{eq:an nonhom}). 

Figure \ref{fig:gaussian} displays this numerical solution, clearly
indicating that no runaway problem occurs. Note that panel (a) of
this figure simultaneously shows the force $F(T)$ and the resultant
acceleration $A(T)$ (the latter actually comes in three variants
as we shortly detail); but these two quantities are indistinguishable
in this graph because the radiation-reaction magnitude is very small,
$\varepsilon=10^{-3}$. To clearly see the radiation-reaction effect
we need to look at the \emph{difference} between $A$ and $F$. To
this end, panel (b) displays the quantity $\alpha(T)\equiv A(T)-F(T)$
(which is entirely due to radiation reaction). 

More specifically, the figure presents three different variants of
$A$ (and $\alpha$): our numerically-constructed solution $A_{N}(T)$,
and the two analytically-constructed solutions used here for comparison,
namely the Dirac-type solution $A_{D}(T)$ and the reduction-of-order
solution $A_{r}(T)$ {[}Eqs. \eqref{eq:dirac gauss} and \eqref{eq:reduced gauss}
respectively{]}. These three variants of $A$ yield three corresponding
variants of $\alpha=A-F$, denoted $\alpha_{N}$, $\alpha_{D}$, and
$\alpha_{r}$ respectively. Panel (b) shows that our numerically-constructed
solution $\alpha_{N}$ is visually-indistinguishable from the two
other, analytically constructed, variants $\alpha_{D}$ and $\alpha_{r}$
(the difference between the three variants is presented below in Fig.
\ref{fig:gaussian diff}). This indicates that our finite-difference
scheme accurately captures the effect of the radiation-reaction force
on the acceleration, despite the small value of $\varepsilon$ used
here. 

Figure \ref{fig:gaussian diff} quantifies the differences between
the three variants of $\alpha$. It displays $\alpha_{r}-\alpha_{D}$
(marked by the dashed purple curve). It also displays $\alpha_{N}-\alpha_{D}$
for three different values of $\delta T$ (used in the construction
of $\alpha_{N}$): $\delta T_{1}=10^{-2}$, $\delta T_{2}=5\cdot10^{-3}$,
and $\delta T_{3}=2.5\cdot10^{-3}$. It indicates first-order convergence
of the finite-difference scheme, as expected. Notice that the smallest
numerical time step used here ($\delta T_{3}$) is only slightly larger
than the minimal allowed value ($2\varepsilon=2\cdot10^{-3}$), but
the numerical scheme is still stable as $|p|=2/3<1$. For this value
of $\delta T$, the ``numerical error'' indicator (namely the deviation
of $\alpha_{N}$ from $\alpha_{D}$ \footnote{It is sensible to use $\alpha_{D}$ as a reference for assessing the
accuracy of $\alpha_{N}$, because $A_{D}(T)$ is an exact closed-form
solution of the EOM.}) is comparable to the difference $\alpha_{r}-\alpha_{D}$.

\begin{figure}
\centering{}\includegraphics[scale=0.4]{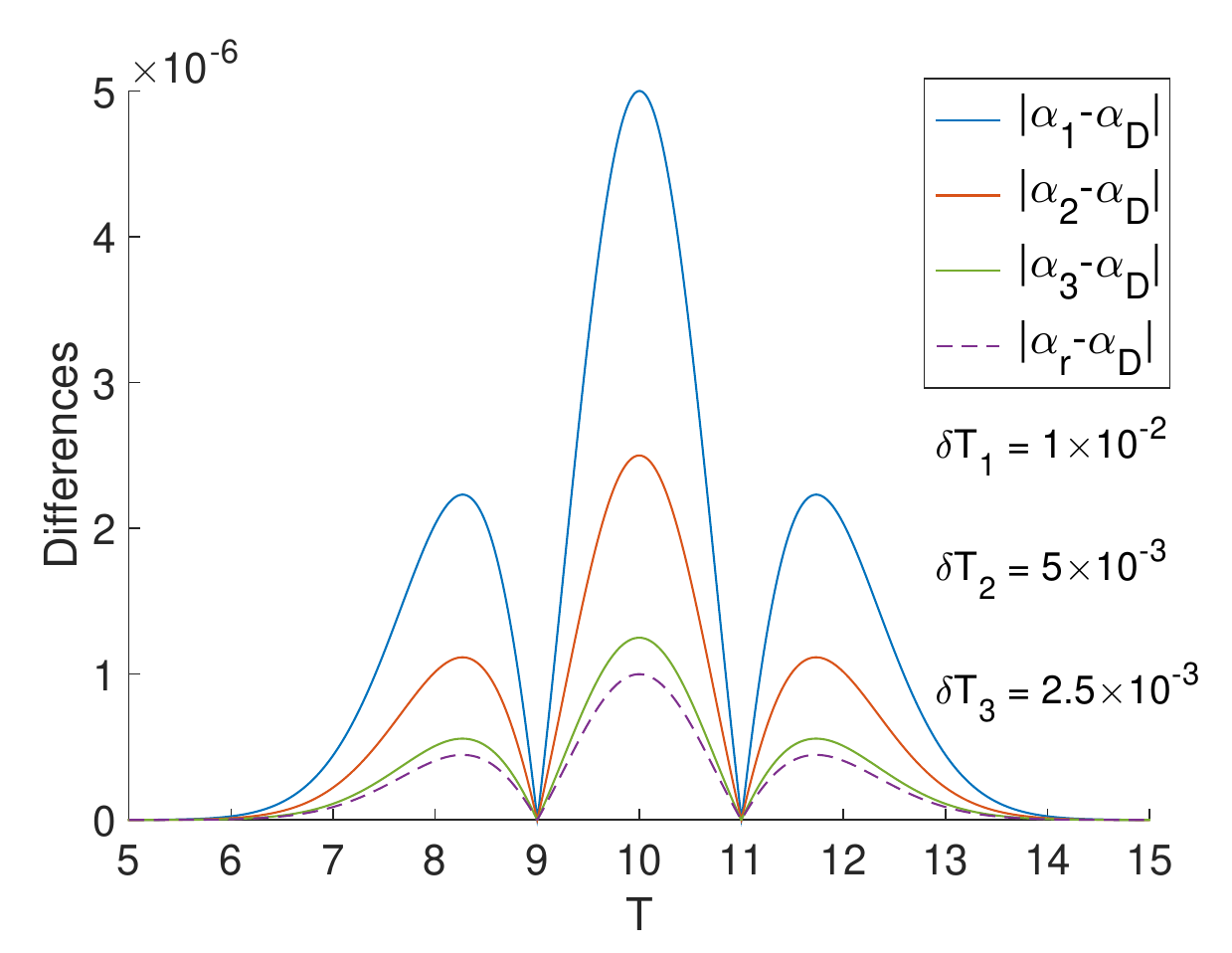}\caption{The differences between the three different variants of $\alpha\left(T\right)$,
for the same Gaussian force as in Fig. \ref{fig:gaussian}a. The dashed
purple curve denotes $\left|\alpha_{r}-\alpha_{D}\right|$. The three
solid curves all denote $\left|\alpha_{N}-\alpha_{D}\right|$ but
for numerically constructed functions $\alpha_{N}(T)$ obtained with
three different time steps: $\delta T_{1}=10^{-2}$ (blue), $\delta T_{2}=5\cdot10^{-3}$
(orange), and $\delta T_{3}=2.5\cdot10^{-3}$ (green). In the legend
we use the abbreviated notation $\alpha_{i}\equiv\alpha_{N}(\delta T=\delta T_{i})$.
 \label{fig:gaussian diff}}
\end{figure}

The analysis in Sec. \ref{sec:Discretizing} indicated that for stable
numerical integration we must choose $\delta T>2\varepsilon$ (to
ensure $|p|<1$). We checked this criterion numerically: For $\delta T=2.01\varepsilon$
the numerics worked very well, but for $\delta T=1.99\varepsilon$
we encountered a severe instability, expressed in an exponentially-growing
error. This was found to be the case in both the Gaussian and the
Sin\textasciicircum{}4 cases.

\subsection{Sin\textasciicircum{}4 External Force}

\begin{figure}
\begin{centering}
\subfloat[]{\centering{}\includegraphics[scale=0.4]{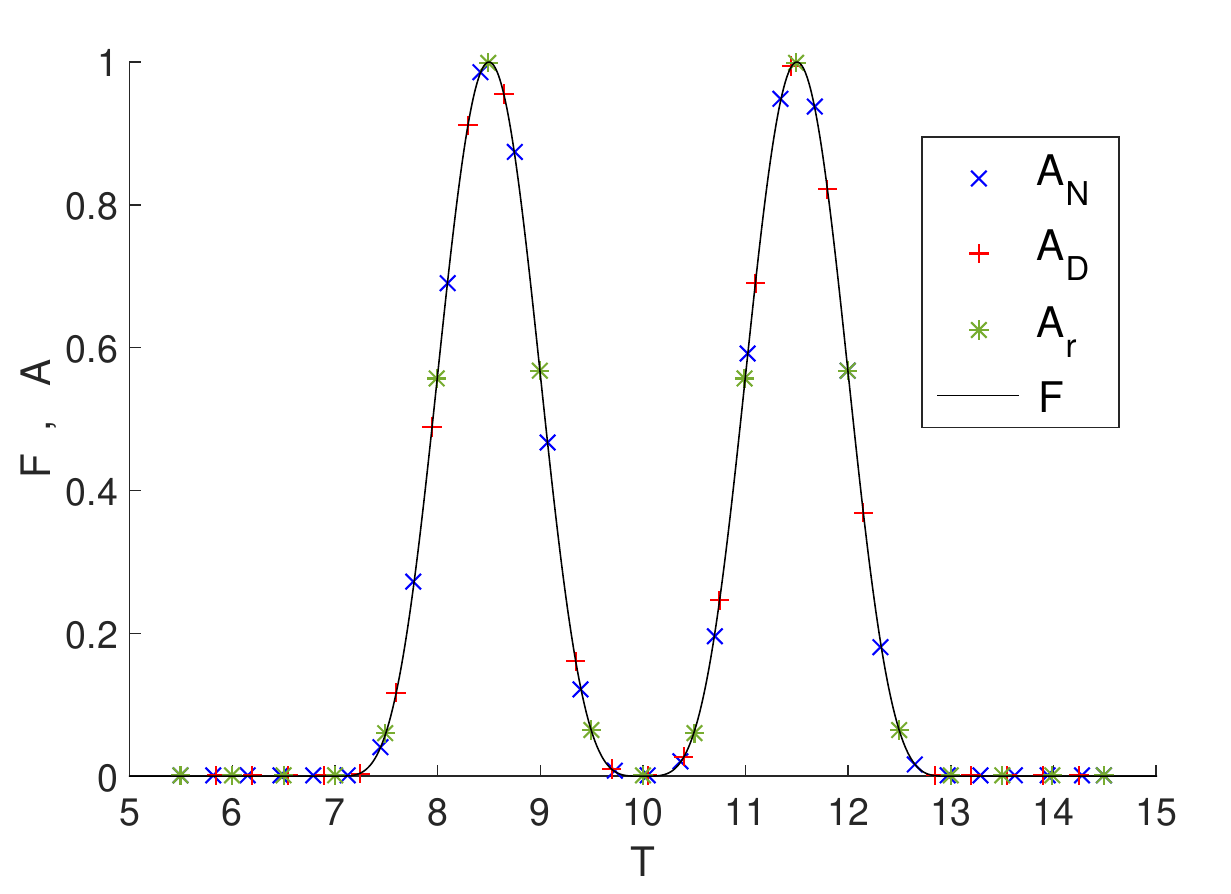}}\subfloat[]{\begin{centering}
\includegraphics[scale=0.4]{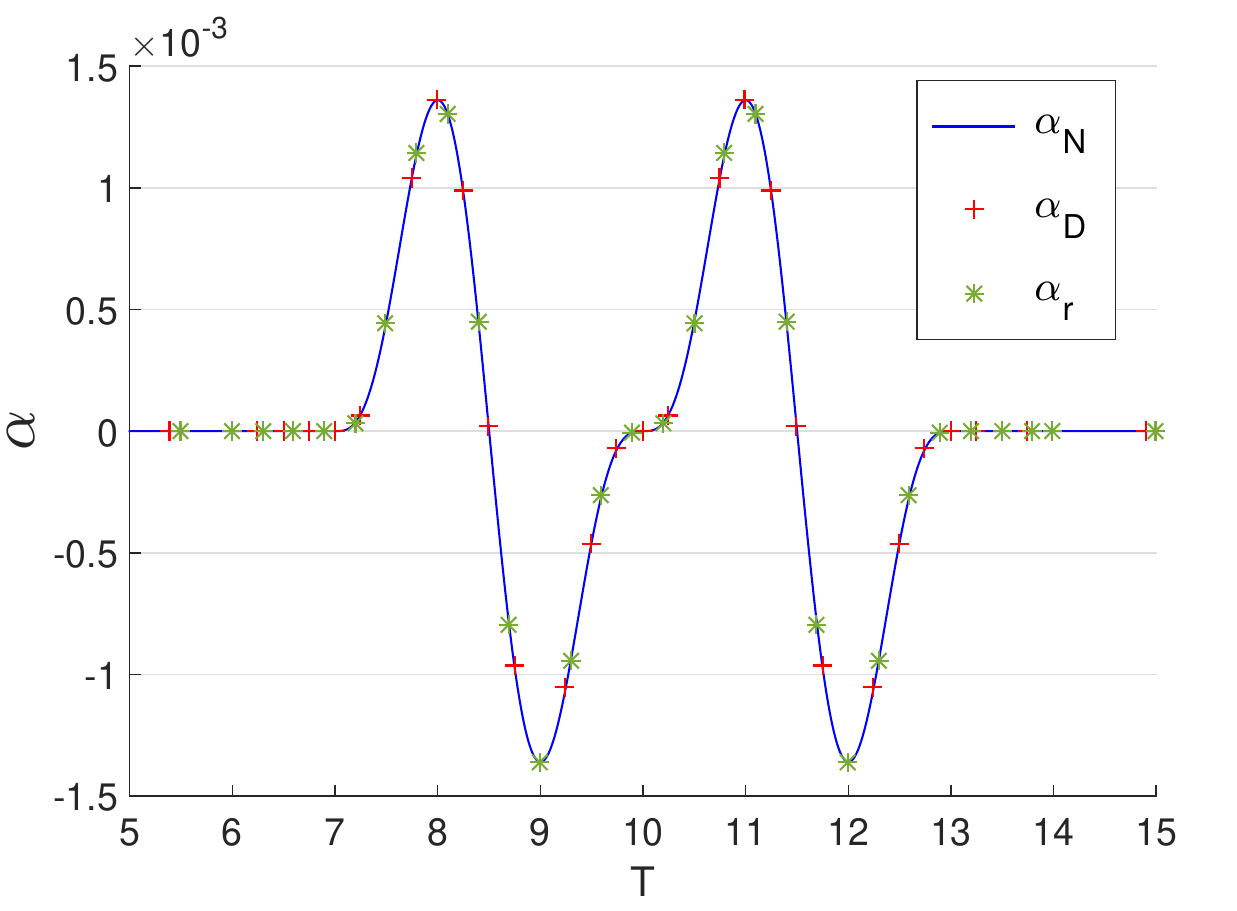}
\par\end{centering}
}
\par\end{centering}
\caption{Sin\textasciicircum{}4 external force. The parameters chosen are $B=1$,
$W=3$, $\mu=10$, and again we use $\varepsilon=10^{-3}$. Presentation
and notation is the same as in Fig. \ref{fig:gaussian} above. In
particular, panel (a) displays both $F(T)$ and $A(T)$ (indistinguishable),
and panel (b) displays $\alpha(T)=A(T)-F(T)$; and as before both
$A$ and $\alpha$ come with their three variants, denoted by the
indices ``$N$'', ``$D$'', or ``$r$''. All symbols are the
same as in the corresponding panels in Fig. \ref{fig:gaussian}. \label{fig:cos}}
\end{figure}

Next we choose the external force $F(T)$ to be the following function:
\begin{equation}
F(T)=B\sin^{4}\left[\frac{\pi}{W}(T-\mu)\right]\Theta(T-\mu+W)\,\Theta(\mu+W-T)\,.\label{eq:cosine force}
\end{equation}
It represents a full period ($T$ interval equals $2W$) of a Sin\textasciicircum{}4
function centered at $T=\mu$, with amplitude $B$. This function,
displayed in Fig. \ref{fig:cos}a, was chosen as an example of a three-times
differentiable function with a compact support. The corresponding
Dirac-type solution (\ref{eq:Dirac General}) is then\footnote{The integral appearing in Eq. \eqref{eq:dirac cos} is elementary
and straightforward, yet the explicit expression is rather lengthy
(and non-illuminating), therefore we do not present it here.} 

\begin{equation}
A_{D}(T)=-\frac{1}{\varepsilon}e^{T/\varepsilon}\int_{\infty}^{T}e^{-s/\varepsilon}B\sin^{4}\left[\frac{\pi}{W}(s-\mu)\right]\Theta(s-\mu+W)\,\Theta(\mu+W-s)ds\,.\label{eq:dirac cos}
\end{equation}
The reduction-of-order solution (\ref{eq:reduction}) now becomes

\begin{equation}
A_{r}(T)=B\,\Theta(T-\mu+W)\,\Theta(\mu+W-T)\sin^{4}\left[\frac{\pi}{W}(T-\mu)\right]\left\{ 1+\varepsilon\frac{4\pi}{W}\cot\left[\frac{\pi}{W}(T-\mu)\right]\right\} .\label{eq:reduced cos}
\end{equation}

We choose here the parameters $B=1$, $W=3$ and $\mu=10$. The radiation-reaction
strength parameter is the same as before, $\varepsilon=10^{-3}$.
We again solve Eq. \eqref{eq: nonhom acceleration} numerically using
the finite-difference scheme constructed in Sec. \ref{sec:Discretizing}. 

\begin{figure}[H]
\centering{}\includegraphics[scale=0.4]{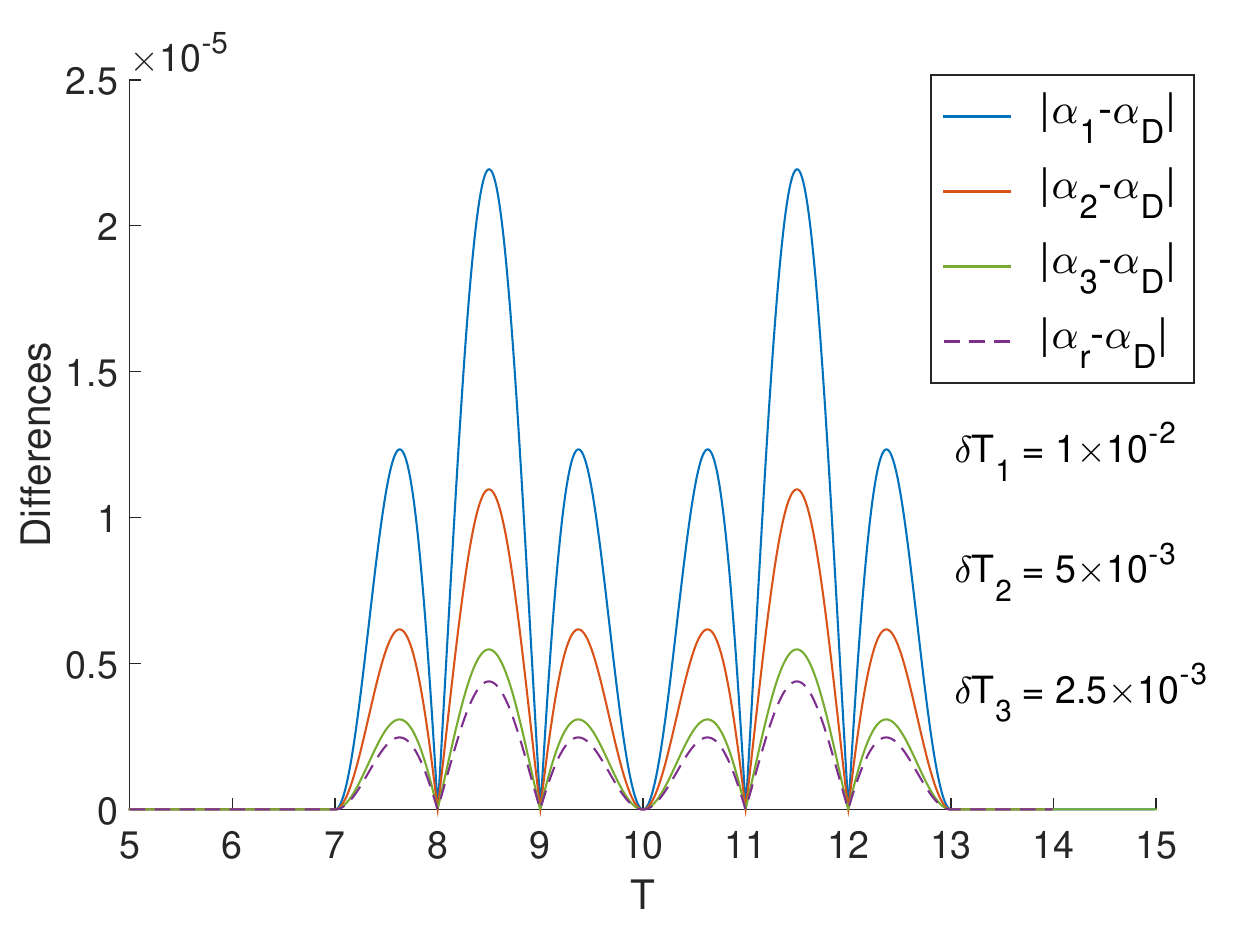}\caption{The differences between the three different variants of $\alpha\left(T\right)$,
for the Sin\textasciicircum{}4 external force. Presentation and notation
is the same as in Fig. \ref{fig:gaussian diff} above. The dashed
purple curve denotes $\left|\alpha_{r}-\alpha_{D}\right|$. The three
solid curves all denote $\left|\alpha_{N}-\alpha_{D}\right|$, for
numerically constructed functions $\alpha_{N}(T)$ obtained with the
same set of three time steps used above: $\delta T_{1}=10^{-2}$ (blue),
$\delta T_{2}=5\cdot10^{-3}$ (orange), and $\delta T_{3}=2.5\cdot10^{-3}$
(green). In the legend we again use the abbreviated notation $\alpha_{i}\equiv\alpha_{N}(\delta T=\delta T_{i})$.\label{fig:cos diff}}
\end{figure}

The results closely resemble the case of Gaussian force presented
above. Figure \ref{fig:cos}a (fully analogous to Fig. \ref{fig:gaussian}a)
displays both $F(T)$ and $A(T)$. The latter again comes in three
variants: our numerical solution $A_{N}(T)$, the Dirac-type solution
$A_{D}(T)$, and the reduction-of-order solution $A_{r}(T)$. Figure
\ref{fig:cos}b (fully analogous to Fig. \ref{fig:gaussian}b) depicts
the back-reaction effect as represented by the difference $\alpha(T)=A(T)-F(T)$
\textemdash{} with its three variants $\alpha_{N}$, $\alpha_{D}$,
and $\alpha_{r}$. Figure \ref{fig:cos diff} (analogous to Fig. \ref{fig:gaussian diff})
shows the differences between these three variants of $\alpha$. In
particular, it displays $\alpha_{N}-\alpha_{D}$ for three different
values of $\delta T$ (the same values $\delta T_{1,2,3}$ used above).
Overall, these graphs lead to the same conclusions as in the case
of a Gaussian force: The runaway problem has been entirely cured in
the numerical integration, the truncation errors associated with the
three time steps $\delta T_{1,2,3}$ are compatible with first-order
convergence, and overall we achieve a good accuracy. Again, for the
smallest time step $\delta T_{3}$ the difference $\alpha_{N}-\alpha_{D}$
is comparable to $\alpha_{r}-\alpha_{D}$. 


\section{Conclusions\label{sec:Conclusions}}

In this paper we have presented a pragmatic method to numerically
construct the physically meaningful solutions of the self-force equation
of motion, thereby overcoming the runaway problem plaguing the analysis
of the motion of a charged point-like mass under the influence of
radiation reaction. This method exploits the fundamental separation
of time scales which usually arises in this problem, by choosing the
numerical time-step to be large compared to the runaway time-scale,
yet much smaller than the characteristic time scale $\tau$ of the
external force (which is also the time-scale of the actual physical
orbit). The finite-difference scheme was designed to accurately incorporate
the effect of the radiation-reaction force at the physically relevant
time scale $\tau$, while completely suppressing the undesired runaway
mode. As a demonstration of this finite-difference method, we have
numerically computed the time-dependent acceleration of a charged
point-like particle subject to two different types of external force
$F_{ext}(t)$, one is Gaussian and the other has the form of Sin\textasciicircum{}4.
The resulting accelerations accurately match the physically expected
ones.

As discussed above, there is a close analogy between the runaway problem
associated with electromagnetic radiation reaction, analyzed in this
paper, and the one that occurs in semiclassical general relativity,
for example in the context of black-hole evaporation. However, the
latter turns out to be significantly more involved, as the semiclassical
Einstein field equation is a PDE rather than ODE \footnote{See also footnote \ref{fn:RSET} above.}.
Still, we expect our finite-difference method to be applicable to
this case as well, despite its technically more challenging nature.

The numerical procedure presented here is basically a first-order
finite-difference scheme, implying that the numerical error scales
like the time step $\delta T$. It may be useful to upgrade the scheme
to second-order accuracy. In this regard we point out that there is
no much point in increasing the numerical accuracy: In the present
scheme the truncation error (e.g. the relative error in $\alpha=A-F$)
is already at order $\sim\varepsilon=\theta/\tau$; and as explained
in the Introduction this is the best one can hope to achieve due to
the inherent limited accuracy in the self-force formula (\ref{eq:AL-force}),
which originates from finite-size effects. Nevertheless, a second-order
numerical scheme will allow increasing $\delta T$, which will in
turn significantly shorten computation times.

In the published version of this paper \cite{next paper}, which includes
a significant extension of this manuscript by A. Lanir and O. Sela,
we extend this investigation to an external force depending on space
as well as time. In that case, one obtains a third-order radiation-reaction
EOM in terms of the position (rather than a first-order equation for
the acceleration). We also extend the analysis further to include
higher order EOMs. We show that our finite-difference method (when
properly generalized) is applicable in this more general framework
as well. 

Subsequently, the method should be extended to the application on
a PDE system with artificial radiation-reaction-like term, in a preparatory
step before applying it to the semiclassical Einstein field equation.

\end{document}